\documentclass{IEEEtran}
\IEEEoverridecommandlockouts
\usepackage{cite}
\usepackage{amsmath,amssymb,amsfonts}
\usepackage{algorithm,algorithmic}
\usepackage{graphicx}
\usepackage{bm}
\usepackage{textcomp}
\usepackage{xcolor}
\def\BibTeX{{\rm B\kern-.05em{\sc i\kern-.025em b}\kern-.08em
    T\kern-.1667em\lower.7ex\hbox{E}\kern-.125emX}}
\begin{document}

\title{MF-based Dimension Reduction Signal Compression for Fronthaul-Constrained Distributed MIMO C-RAN}

\author{\large \IEEEauthorblockN{Fred Wiffen\IEEEauthorrefmark{1}, Mohammud Z. Bocus\IEEEauthorrefmark{2}, Angela Doufexi\IEEEauthorrefmark{1} and Woon Hau Chin\IEEEauthorrefmark{2}} \\
\IEEEauthorblockA{\IEEEauthorrefmark{1}Communication Systems \& Networks Research Group, University of Bristol,
Bristol, UK}
 \IEEEauthorblockA{\IEEEauthorrefmark{2}Toshiba Research Europe Limited, Bristol, UK}
\IEEEauthorblockA{Email: fred.wiffen@bristol.ac.uk}
}

\maketitle

\begin{abstract}
In this work we propose a fronthaul compression scheme for distributed MIMO systems with multi-antenna receivers, in which, prior to signal quantisation, dimension reduction is performed at each receiver by matched filtering the received signal with a subset of the local user channel vectors. By choosing these matched filter vectors based on global channel information, a high proportion of the potential capacity may be captured by a small number of signal components, which can then be compressed efficiently using local signal compression. We outline a greedy algorithm for selecting the matched filtering vectors for each receiver, and a local transform coding approach for quantising them, giving expressions for the resulting system sum and user capacities. We then show that the scheme is easily modified to account for imperfect CSI at the receivers. Numerical results show that with a low signal dimension the scheme is able to operate very close to the cut-set bound in the fronthaul-limited regime, and demonstrates significant improvements in rate-capacity trade-off versus local compression at all operating points, particularly at high SNR.

\end{abstract}

\begin{IEEEkeywords}
distributed MIMO, C-RAN, fronthaul compression, dimension reduction
\end{IEEEkeywords}

\section{Introduction}
In a distributed multiple input multiple output (MIMO) uplink system, $K$ users are jointly served by $L$ receivers (or remote radio heads), each equipped with $M$ antennas and distributed geographically within the service area. This distribution of antennas provides macro-diversity and improves uniformity of service, and is facilitated by the recent shift towards a cloud radio-access-network (C-RAN) architecture, in which processing for multiple receivers is performed at a single central processor (CP). A significant practical challenge with C-RAN MIMO, however, is the transfer of data from the receivers to the CP -- the large data rates associated with the transfer of raw IQ samples \cite{brubaker2016emerging}, combined with a growing interest in replacing fixed fibre with reduced capacity wireless point-to-point connections \cite{lombardi2018microwave} resulting in a need for efficient lossy compression of the received signals.

The challenge of data compression for fronthaul constrained MIMO networks has received much research attention, see e.g. \cite{7444125} and references therein. We restrict our attention to the uplink compress-and-forward architecture, in which compression is applied at the receivers before forwarding to the CP for global symbol detection. For simplicity, schemes in which each receiver independently compresses and forwards its own signal, e.g. \cite{6226311}, are attractive and currently implemented in practical systems, but do not exploit the inherent dependencies between signals at different receivers and therefore do not efficiently make use of the available fronthaul. On the other hand, the best performance is achieved through the use of distributed source coding techniques, in which signals at all receivers are jointly compressed and decoded, for example \cite{6824778}, but these have increased computational complexity. Point-to-point compression schemes in which signals at each receiver are separately compressed and decoded, but using compression codebooks that are jointly designed to exploit dependencies between receivers, present an attractive compromise. 

The optimal point-to-point compression scheme for the Gaussian channel involves solving a non-convex optimization to find the quantisation noise covariances for each receiver, using, for example, a successive convex approximation approach \cite{7063645}. However, this method does not scale well to large networks with rapidly changing channels. Sub-optimal approaches include \cite{7134796} and \cite{8762078}, which apply transform coding, with a local decorrelating transform applied to the signals, followed by a centrally performed rate allocation stage. An interesting observation in \cite{7134796} is that at lower fronthaul rates the optimal rate allocation is sparse - only a subset of available signal components at each receiver are quantised. The scheme can thus be seen as effectively performing dimension reduction on the received signal. Signal dimension reduction is also employed in \cite{8671721} using an analog beamforming stage, which is followed by a digital compression stage. The dimension reduction concept has parallels to the downlink sparse beamforming approach \cite{6920005}, in which each transmitter only transmits to a subset of the users, reducing the number of data streams that need to be transferred over fronthaul.

\subsection*{Paper Overview}
In this paper we focus on systems with single antenna users, and an overall excess of receive antennas, $ML \gg K$. We propose a fully-digital dimension reduction based compression scheme, in which each receiver reduces its signal dimension by filtering its signal in the direction of a subset of $N<K$ users before applying local compression to the signals and forwarding them to the CP. The key feature of this scheme is that the lossy dimension reduction stage produces a reduced number of signal components with reduced inter-receiver dependencies, such that local signal compression can be applied efficiently. 

The paper makes the follow contributions:
\begin{itemize}
    \item a greedy algorithm for selecting the dimension reduction MF vectors is proposed.
    \item a transform coding compression algorithm is outlined.
    \item the scheme is adapted for the case of imperfect CSI.
    \item capacity equations are provided for both optimal and linear symbol detection.
    \item numerical results for Rayleigh fading channels are given, showing that:
    \begin{itemize}
        \item a significantly reduced signal dimension can be used at each receiver whilst only losing a small proportion of the total information captured by the full dimension signal.
        \item the scheme significantly outperforms local compression at all fronthaul rates, especially at high SNR.
        \item with a low signal dimension, the proposed scheme can operate very close to the cut-set upper bound in the fronthaul-limited region.
        \item good performance is also achieved under linear symbol detection, and with imperfect CSI at the receivers.
\end{itemize}
\end{itemize}
The paper is structured as follows: Section \ref{sec:systemmodel} outlines the system model and configurations used for numerical examples, with Section \ref{sec:rdcomp} providing a rationale for dimension reduction and outlining the overall scheme. Section \ref{sec:mfdr} outlines the greedy dimension reduction algorithm and gives insights into its behaviour before Section \ref{sec:comp} outlines the transform coding compression scheme. Section \ref{sec:cap} gives capacity equations. Section \ref{sec:CSI} adapts the dimension reduction compression scheme for the case of imperfect CSI and Section \ref{sec:practical} recommends some modifications to the system for practical implementation. Finally, Section \ref{sec:numerical} provides numerical results.

\section{System Model}
\label{sec:systemmodel}
\subsection{System Model}
We consider an uplink system in which $L$ distributed MIMO receivers, each equipped with $M$ antennas, jointly serve $K$ single antenna users, where there is an overall excess of receive antennas, $ML \gg K$. The receivers have digital processing capability and are connected via individual fronthaul links with capacity $\mathcal{R}$ bits per channel use (bpcu) to a central processor (CP), which uses signals from all of the receivers to jointly detect and decode the transmitted user symbols. 

\begin{figure}
    \centering
    \includegraphics[width=0.95\linewidth]{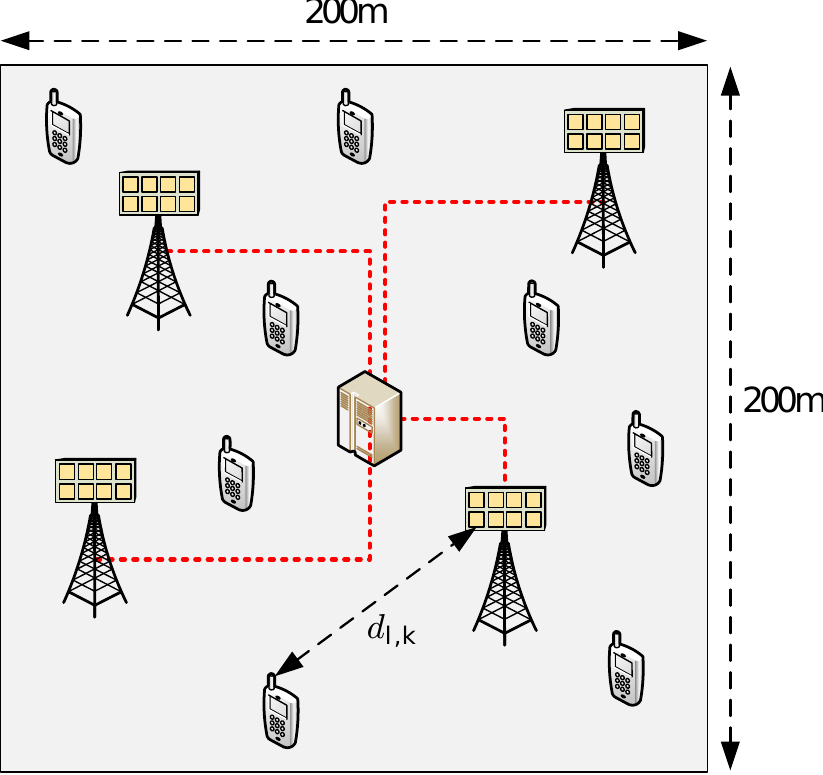}
    \caption{Illustration of system configuration with $K=8$, $L=4$, $M = 8$.}
    \label{fig:simconfig}
\end{figure}

The received uplink signal at receiver $l$ is given by
\begin{equation}
    \mathbf{y}_l = \mathbf{H}_l\mathbf{x} + \bm{\eta},
\end{equation}
where $\mathbf{H}_l \in \mathbb{C}^{M \times K}$ is the channel to receiver $l$, $\bm{\eta}$ additive white Gaussian noise with unit variance
\begin{equation}
    \bm{\eta} \sim \mathcal{CN}\big(0,\mathbf{I}_{M}\big).
\end{equation}
and $\mathbf{x}$ independent Gaussian uplink symbols with signal-to-noise ratio (SNR) $\rho$
\begin{equation}
    \mathbf{x} \sim \mathcal{CN}\big(0,\rho\mathbf{I}_{K}\big).
\end{equation}
Column $k$ of $\mathbf{H}_l$ is the channel vector between user $k$ and receiver $l$
\begin{equation}
    \mathbf{H}_l = \begin{bmatrix} \mathbf{h}_{l,1} & \ldots & \mathbf{h}_{l,K} \end{bmatrix},
\end{equation}
where we assume that the $\mathbf{H}_l$ are full rank, i.e. $t = \min(M,K)$. Each receiver has access to local CSI, and the CP has access to CSI as applicable.

\subsection{Numerical Example Configurations}
Illustrative numerical examples provided throughout this paper consider a single cell with the users and receivers positioned randomly within a 200m $\times$ 200m area, with user height 1 m and receiver height 6 m. The user channels follow complex normal independent fading 
\begin{equation}
    \mathbf{h}_{l,k} \sim \mathcal{CN}\big(0,p_k\beta_{l,k}\mathbf{I}_M\big)
\end{equation}
where $p_k$ is the uplink power control coefficient for user $k$, and $\beta_{l,k}$ follows a log-distance path loss model with path loss exponent 2.9 and shadow fading 5.7 dB \cite{7434656}. Power control is applied such that the total average received power for each user is the same
\begin{equation}
    \frac{1}{ML}\mathbb{E}\big[\sum_l \big \Vert \mathbf{h}_{l,k} \big \Vert^2  \big] =  p_k\sum_{l=1}^L\frac{\beta_{l,k}}{L} = 1\nonumber.
\end{equation}
All mean quantities are averaged over both channel realisation and user \& receiver locations. Note that the general methods outlined in this paper do not rely on any channel model assumptions.

\section{Dimension Reduction for Signal Compression}
\label{sec:rdcomp}
Consider local signal compression, where the compressed signal, $\tilde{\mathbf{y}}_l$, is chosen to maximise the information it provides about the user symbols
\begin{equation}
\begin{aligned}
& \underset{\tilde{\mathbf{y}}_l}{\text{maximise}} \quad \mathcal{I}\big(\tilde{\mathbf{y}}_l;\mathbf{x}\big)  \\
& \mathrm{subject \ to} \quad \mathcal{I}\big(\tilde{\mathbf{y}}_l;\mathbf{y}_l\big) \leq \mathcal{R}.
\end{aligned}
\end{equation}
In \cite{6875354} it is shown that this is achieved by a transform coding approach in which a local decorrelating transform is applied at each receiver to produce $t$ signal components, which are quantised using $t$ scalar quantisers with appropriate local rate allocation. The resulting quantisation noise power for each signal component decreases approximately exponentially (see Section \ref{sec:comp}) with $\mathcal{R}/t$ 
\begin{equation}
    \mathrm{quantisation \ noise} \sim 2^{-\mathcal{R}/t}.
\end{equation}
However, in a distributed MIMO system, the received signals are inherently correlated through their dependence on $\mathbf{x}$, and local compression performs poorly. If at each receiver we take a reduced number $N<t$ of signal components, and apply local compression, the quantisation noise can be reduced,
\begin{equation}
    \mathrm{quantisation \ noise} \sim 2^{-\mathcal{R}/N}.
\end{equation}
Clearly, this dimension reduction causes a loss of system capacity. However, this is not inherently problematic, since any signal compression process necessarily entails a capacity loss. If the signal components are chosen at a \textit{global} level to account for inter-receiver signal dependencies, the information loss due to dimension reduction can be kept small, whilst overall system capacity increased due to the reduction in quantisation noise.

The physical distribution of users and receivers means that if $N$ signal components are chosen at one receiver that provides a lot of information about the signals of $N$ users to which it has strong channels, the additional information about those users that receivers with weaker channels to them can provide is small. By appropriately choosing the signal component we can therefore expect that a reduced number of signal dimensions can capture a high proportion of the channel capacity.

A more in-depth look at the reduced dimension signal compression concept is provided in \cite{wiffen2020dimension}.

\subsection*{Proposed Scheme}
Based on the insights above, a good approach is for each receiver to filter its received signal in the direction of a subset of $N<t$ of the users. This can be achieved by matched filtering using channel vectors associated with a subset of users,
\begin{equation}
    \mathbf{z}_l = \mathbf{F}_l^\dagger\mathbf{y}_l,
\end{equation}
where the $N$ columns of $\mathbf{F}_l$ are the channel vectors for the subset of selected users, $\mathcal{S}_l$, at receiver $l$,
\begin{equation}
    \mathbf{F}_{l} = \begin{bmatrix} \mathbf{h}_{l,\mathcal{S}_l(1)} & \ldots & \mathbf{h}_{l,\mathcal{S}_l(N)}\end{bmatrix}.
\end{equation}
We pick the MF vectors for each receiver at the CP using global CSI, to maximise the joint mutual information provided by the reduced dimension signals
\begin{equation}
    \underset{\mathcal{S}_1,\ldots,\mathcal{S}_L}{\mathrm{maximise}} \quad  \mathcal{I}\big(\mathbf{z}_1,\ldots,\mathbf{z}_L ; \mathbf{x}\big).
\end{equation}
The receivers then perform local signal compression on the reduced dimension signals for transfer to the CP,
\begin{equation}
\begin{aligned}
& \underset{\tilde{\mathbf{z}}_l}{\text{maximise}} \quad \mathcal{I}\big(\tilde{\mathbf{z}}_l;\mathbf{x}\big)  \\
& \mathrm{subject \ to} \quad \mathcal{I}\big(\tilde{\mathbf{z}}_l;\mathbf{z}_l\big) \leq \mathcal{R}.
\end{aligned}
\end{equation}
The sum capacity of the distributed MIMO system is given by
\begin{equation}
\label{eqn:sumcap2}
    \mathcal{C}_{\mathrm{sum}} = \mathcal{I}\big(\tilde{\mathbf{z}}_1,\ldots,\tilde{\mathbf{z}}_L;\mathbf{x}\big).
\end{equation}

\section{MF-Based Dimension Reduction}
\label{sec:mfdr}
Each receiver applies a dimension reduction filter to its received signal
\begin{equation}
    \mathbf{z}_l = \mathbf{F}_l^\dagger\mathbf{y}_l.
\end{equation}
Using the QR decomposition this filter may be written
\begin{equation}
    \mathbf{F}_{l} = \mathbf{Q}_l\mathbf{R}_l
\end{equation}
where $\mathbf{Q}_l \in \mathbb{C}^{M \times N}$ has orthonormal columns, $\mathbf{q}_{l,i}$, and $\mathbf{R}_l \in \mathbb{C}^{N \times N}$ is upper triangular. If the columns of $\mathbf{F}_l$ are linearly independent, $\mathbf{R}_l$ is invertible, and therefore by the data processing inequality 
\begin{equation}
    \mathcal{I}\big(\mathbf{z}_1,\ldots,\mathbf{z}_L;\mathbf{x}\big) = \mathcal{I}\big(\bar{\mathbf{z}}_1,\ldots,\bar{\mathbf{z}}_L;\mathbf{x}\big)
\end{equation}
where
\begin{equation}
\begin{aligned}
    \bar{\mathbf{z}}_l &= \mathbf{Q}_l^\dagger\mathbf{y}_l = \mathbf{Q}_l^\dagger\mathbf{H}_l\mathbf{x} + \bar{\boldsymbol{\eta}}
\end{aligned}
\end{equation}
i.e. the information in the filtered signal depends only on the $N$ dimensional-subspace spanned by the selected user vectors.
The columns of $\mathbf{Q}_l$ may be calculated iteratively using the Gram-Schmidt procedure
\begin{equation}
\label{eqn:GSvectors}
    \mathbf{q}_{l,i} = \frac{\mathbf{P}_{l,i}\mathbf{h}_{l,\mathcal{S}_l(i)}}{\Vert \mathbf{P}_{l,i}\mathbf{h}_{l,\mathcal{S}_l(i)} \Vert}
\end{equation}
where
\begin{equation}
    \mathbf{P}_{l,i} = \mathbf{I}_M - \sum_{j<i} \mathbf{q}_{l,j}\mathbf{q}_{l,j}^\dagger
\end{equation}
The joint mutual information is 
\begin{equation}
    \mathcal{I}\big(\bar{\mathbf{z}}_1,\ldots,\bar{\mathbf{z}}_L;\mathbf{x}\big) = \log_2\det\big(\mathbf{I}_K + \rho\sum_{l=1}^L\sum_{i=1}^N\mathbf{H}_l^\dagger\mathbf{q}_{l,i}\mathbf{q}_{l,i}^\dagger\mathbf{H}_l\big)
\end{equation}
The problem of selecting the optimal set of users vectors for all receivers is combinatorial with ${K \choose N}^L$ possible combinations, and hence an exhaustive search is prohibitive. A more tractable approach is to use a greedy algorithm to select the user vectors one at a time, such that each selection stage maximises the mutual information. 

\subsection{Greedy Algorithm}
If after $n$ stages the set of selected MF vectors at receiver $l$ is $\mathcal{S}_l^{(n)}$, the joint mutual information is
\begin{equation}
    \log_2\det\big(\mathbf{I}_K + \rho\sum_{l=1}^L\sum_{i =1} ^{\vert\mathcal{S}_l^{(n)}\vert}\mathbf{H}_l^\dagger\mathbf{q}_{l,i}\mathbf{q}_{l,i}^\dagger\mathbf{H}_l\big),
\end{equation}
which may be written using the matrix determinant lemma
\begin{equation}
     \log_2\det\big(\mathbf{A}_{n-1}^{-1}\big) + \log_2\big(1 + \rho\mathbf{q}_{l,i}^\dagger\mathbf{H}_l\mathbf{A}_{n-1}\mathbf{H}_l^\dagger\mathbf{q}_{l,i}\big)
\end{equation}
where
\begin{equation}
    \mathbf{A}_{n-1} = \big(\mathbf{I}_K + \rho\sum_{l=1}^L\sum_{i =1} ^{\vert\mathcal{S}_l^{(n-1)}\vert}\mathbf{H}_l^\dagger\mathbf{q}_{l,i}\mathbf{q}_{l,i}^\dagger\mathbf{H}_l\big)^{-1}
\end{equation}
Substituting (\ref{eqn:GSvectors}), the information at stage $n$ is maximised by choosing the user vector at receiver $l$ that maximises
\begin{equation}
    \max_{k \notin \mathcal{S}_l^{(n-1)}} \quad \frac{\mathbf{h}_{l,k}^\dagger\mathbf{P}_{l,i}\mathbf{H}_l\mathbf{A}_{n-1}\mathbf{H}_l^\dagger\mathbf{P}_{l,i}\mathbf{h}_{l,k}}{\Vert\mathbf{P}_{l,i}\mathbf{h}_{l,k}\Vert^2}.
\end{equation}
The $\mathbf{A}_{n-1}$ matrix can then be updated using a rank-1 update
\begin{equation}
    \mathbf{A}_{n} = \mathbf{A}_{n-1} - \dfrac{\mathbf{A}_{n-1}\mathbf{H}_{l}^\dagger\mathbf{q}_{l,i}\mathbf{q}_{l,i}^{\dagger}\mathbf{H}_l\mathbf{A}_{n-1}}{1/\rho + \mathbf{q}_{l,i}^{\dagger}\mathbf{H}_l^\dagger\mathbf{A}_{n-1}\mathbf{H}_l\mathbf{q}_{l,i}}.
\end{equation}
This greedy selection can be carried out in a round-robin manner, selecting a MF vector for each receiver in turn, as shown in Algorithm \ref{alg:MFGS}. We refer to this as the matched-filter Gram-Schmidt (MF-GS) algorithm.
\begin{algorithm}[h]
\begin{algorithmic}
\caption{MF-GS Algorithm}
\label{alg:MFGS}
\STATE \textbf{inputs:} $\mathbf{H}_l \quad \forall l$
\vspace{5px}
\STATE $\mathbf{A} \gets \mathbf{I}_K$ \\\vspace{3px}
\STATE $\mathbf{P}_{l} \gets \mathbf{I}_M \quad \forall l$ \hfill \vspace{3px} \\
\STATE $\mathcal{S}_l[1:N] \gets 0 \quad \forall l$ \hfill \textit{sets of select user vectors}\\ \vspace{3px}
\FOR {$n = 1 : N$}
    \FOR {$l = 1 : L$}
        \STATE {
        $k' \gets \underset{k \notin \mathcal{S}_l}{\arg \max} \quad \dfrac{\mathbf{h}_{l,k}^\dagger\mathbf{P}_l\mathbf{H}_{l}\mathbf{A}\mathbf{H}_l^\dagger\mathbf{P}_l\mathbf{h}_{l,k}}{\mathbf{h}_{l,k}^\dagger\mathbf{P}_l\mathbf{h}_{l,k}}$ \hfill \textit{select vector} \\\vspace{3px}
        $\mathbf{q} \gets \dfrac{\mathbf{P}_l\mathbf{h}_{l,k'}}{\Vert \mathbf{P}_l\mathbf{h}_{l,k'} \Vert}$ \hfill \textit{store} \\ \vspace{5px}
        $\mathcal{S}_l[n] \gets k'$ \hfill \textit{index of selected user vector}\\ \vspace{3px}
        $\mathbf{A} \gets \mathbf{A} - \dfrac{\mathbf{A}\mathbf{H}_{l}^\dagger\mathbf{q}\mathbf{q}^{\dagger}\mathbf{H}_l\mathbf{A}}{1/\rho + \mathbf{q}^{\dagger}\mathbf{H}_l^\dagger\mathbf{A}\mathbf{H}_l\mathbf{q}}$ \hfill \textit{rank-1 inverse update} \\ \vspace{5px}
        $\mathbf{P}_l \gets \mathbf{P}_l - \mathbf{q}\mathbf{q}^\dagger$ \hfill \textit{update projection matrix} \\ \vspace{5px}
                }
    \ENDFOR
\ENDFOR \vspace{5px}
\STATE \textbf{outputs:} $\mathcal{S}_l$
\end{algorithmic}
\end{algorithm}
For good performance, the number of signal components available to the CP must be at least the number of users, i.e. $N \geq K/L$. 

\subsection{Algorithm Behaviour}
\begin{figure}[h]
    \centering
    \includegraphics[width=0.95\linewidth]{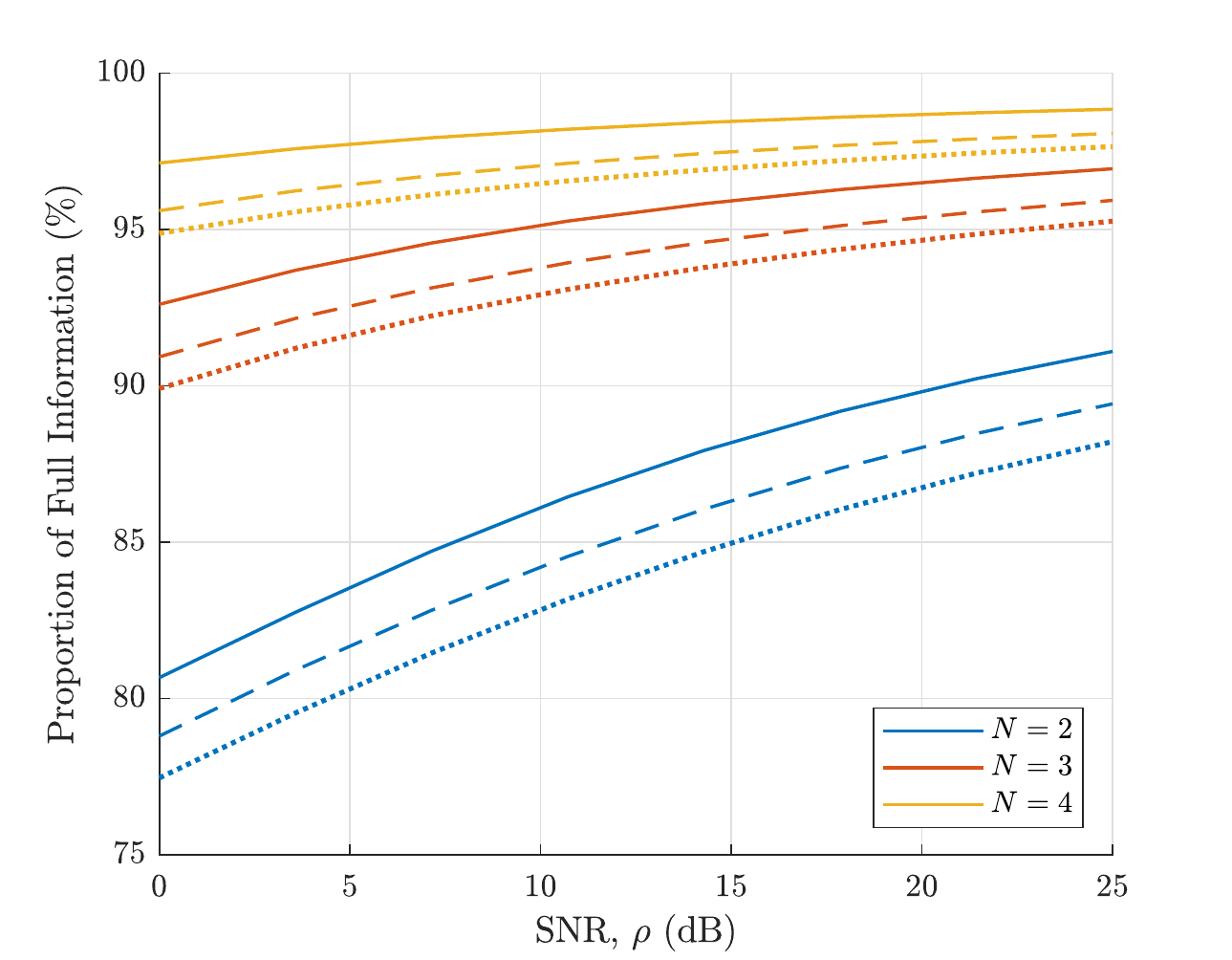}
    \caption{Mean proportion of full dimension mutual information captured by reduced dimension signals, with varying SNR. Solid line: $K=8$, $L=4$, $M = 8$. Dashed line: $K=16$, $L=8$, $M = 8$. Dotted line: $K=24$, $L=12$, $M = 8$.}
    \label{fig:dimredprop}
\end{figure}

We now provide some insights into the behaviour of the MF-GS algorithm. Dropping subscripts for clarity, at each selection stage, the mutual information is increased by
\begin{equation}
    \log_2\big(1 + \rho\mathbf{q}^\dagger\mathbf{H}\mathbf{A}\mathbf{H}^\dagger\mathbf{q}\big)
\end{equation}
where $\mathbf{A}$ can be written using the eigendecomposition
\begin{equation}
    \mathbf{A} = \mathbf{U}\big(\mathbf{I}_K + \rho\mathbf{\Upsilon}\big)^{-1}\mathbf{U}^\dagger
\end{equation}
with $\mathbf{\Upsilon}$ a diagonal matrix containing the $K$ ordered eigenvalues, $\upsilon_i$, of the equivalent channel $\big(\sum_l\sum_i\mathbf{H}_l^\dagger\mathbf{q}_{l,i}\mathbf{q}_{l,i}^\dagger\mathbf{H}_l\big)$, and $\mathbf{U} = \begin{bmatrix} \mathbf{u}_1 & \ldots & \mathbf{u}_K \end{bmatrix}$ the corresponding eigenvectors.
Defining the normalised signal power, $\gamma$, and normalised vector, $\mathbf{c}$, 
\begin{equation}
    \gamma = \mathbf{q}^\dagger\mathbf{H}\mathbf{H}^\dagger\mathbf{q}, \qquad \mathbf{c} = \frac{\mathbf{H}^\dagger\mathbf{q}}{\big\Vert \mathbf{H}^\dagger\mathbf{q} \big\Vert},
\end{equation} 
the mutual information increase can be written
\begin{equation}
\label{eqn:alt}
    \log_2\big(1 + \rho\mathbf{q}^\dagger\mathbf{H}\mathbf{A}\mathbf{H}^\dagger\mathbf{q}\big)
    = \log_2 \Big(1 + \gamma \sum_{i=1}^K \frac{\rho\vert \mathbf{u}_i^\dagger\mathbf{c} \vert^2}{1 + \rho\upsilon_i}\Big),
\end{equation}
where $\mathbf{u}_i^\dagger\mathbf{c}$ is the projection of $\mathbf{c}$ onto eigenvector $i$, with
\begin{equation}
    \sum_{i=1}^K \vert \mathbf{u}_i^\dagger\mathbf{c} \vert^2 = 1.
\end{equation}
From (\ref{eqn:alt}) we can observe that:
\begin{enumerate}
    \item simply selecting the candidate vector that contains the most signal power (large $\gamma$) is a sub-optimal strategy. If the eigenvalue spread of the equivalent channel is large, then signals that project mainly onto the weaker eigenvectors may be selected, despite having lower power. Hence it is not generally optimal to just select the MF vectors corresponding to the $N$ strongest user channels at each receiver. 
    \item for a given signal power, $\gamma$, the best possible candidate signal lies parallel to the eigenvector associated with the smallest eigenvalue, i.e. $\mathbf{c} = \mathbf{u}_K$. This optimal vector can be shown to increase the smallest eigenvalue of the equivalent channel from $\upsilon_K$ to $\upsilon_K + \gamma$. Similarly, the worst possible candidate signal lies parallel to the largest eigenvector, $\mathbf{c} = \mathbf{u}_1$, increasing the largest eigenvalue from $\upsilon_1$ to $\upsilon_1 + \gamma$. Furthermore, it can be shown that any choice of $\mathbf{c}$ gives an updated equivalent channel with all $\upsilon'_i \geq \upsilon_i$.
    \item when the $\upsilon_i$ are large, the information provided by additional signal components reduces, and for $\rho\upsilon_i \gg 1$ is independent of $\rho$.
\end{enumerate}
From these observations we can expect that at each stage the MF-GS algorithm will generally act to make selections that increase the smaller eigenvalues of the equivalent channel, and tend to produce a full rank equivalent channel matrix (all $\upsilon_i > 0$) when $LN \geq K$. As more selections are made and the channel eigenvalues increase, the capacity will grow more slowly with each selection, and hence there are diminishing returns from increasing $N$. For large $\rho$ (and all $\upsilon_i > 0$) the information loss due to dimension reduction is independent of $\rho$, and hence the proportion of information lost vanishes as $\rho \to \infty$, as shown in Figure \ref{fig:dimredprop} (where $t=8$). We see that even with small $N$, a high proportion of the available information can be captured.

\section{Signal Compression}
\label{sec:comp}
The reduced dimension signals are compressed at each receiver separately using locally optimal compression. We may equivalently compress either $\mathbf{z}_l$ or $\bar{\mathbf{z}}_l$, choosing $\bar{\mathbf{z}}_l$ for simplicity of analysis
\begin{equation}
\begin{aligned}
& \underset{\tilde{\mathbf{z}}_l}{\text{maximise}} \quad \mathcal{I}\big(\tilde{\mathbf{z}}_l;\mathbf{x}\big)  \\
& \mathrm{subject \ to} \quad \mathcal{I}\big(\tilde{\mathbf{z}}_l;\bar{\mathbf{z}}_l\big) \leq \mathcal{R}, 
\end{aligned}
\end{equation}
As discussed above, this is achieved by applying a linear decorrelating transform to $\bar{\mathbf{z}}_l$ to produce a set of independent variables which are then independently quantised using $N$ scalar quantisers \cite{6875354},
\begin{equation}
    \tilde{\mathbf{z}}_l = \mathbf{V}_l^\dagger\bar{\mathbf{z}}_l + \boldsymbol{\delta}_l
\end{equation}
where $\boldsymbol{\delta}_l \sim \mathcal{CN}\big(0,\mathbf{\Phi}_l\big)$ is the resulting quantisation noise, with diagonal covariance
\begin{align}
    \big[\mathbf{\Phi}_l\big]_{i,i} &= \frac{\big[\mathbf{V}_l^\dagger\mathbb{E}\big[ \bar{\mathbf{z}}_l\bar{\mathbf{z}}_l^\dagger\big]\mathbf{V}_l\big]_{i,i}}{2^{r_{l,i}} - 1}\\
    &= \frac{\rho\lambda_{l,i} +1}{2^{r_{l,i}} - 1}
\label{eqn:quantnoise}
\end{align}
where $\mathbf{V}_l\mathbf{\Lambda}_l\mathbf{V}_l^\dagger = \mathbf{Q}_l^\dagger\mathbf{H}_l\mathbf{H}_l^\dagger\mathbf{Q}_l$, with $\mathbf{\Lambda}_l = \mathrm{diag}(\lambda_{l,i})$. The optimal rate allocation is given by the waterfilling solution
\begin{equation}
\label{eqn:singleratealloc}
    r_{l,i} = \bigg[ \frac{\mathcal{R}}{N_l} + \log_2( \lambda_{l,i}) - \frac{1}{N_l}\sum_{j=1}^{N_l}\log_2(\lambda_{l,j}) \bigg]^+
\end{equation}
where $[a]^+ =\max(0,a)$ and $N_l$ the corresponding number of $r_{l,i} > 0$. Note that for the full dimension case ($N=t$) the $\lambda_{l,i}$ are the eigenvalues of $\mathbf{H}_l\mathbf{H}_l^\dagger$. Assuming $\mathcal{R}$ is sufficiently large that all $N$ dimensions are quantised ($N_l = N$) substituting (\ref{eqn:singleratealloc}) into (\ref{eqn:quantnoise}) the quantisation noise power is approximately
\begin{align}
    \big[\mathbf{\Phi}_l\big]_{i,i} \approx \rho\Big(\prod_{j=1}^N \lambda_{l,j}\Big)^{1/N} 2^{-\mathcal{R}/N},
\end{align}
which is tight for $\rho\lambda_{l,i} \gg 1$, $2^{\mathcal{R}/N} \gg 1$.

\section{Achievable Rates}
\label{sec:cap}
The combined action of the propagation channel, dimension reduction filter and decorrelating transform can be described by an equivalent channel
\begin{equation}
\begin{aligned}
    \tilde{\mathbf{z}}_l &= \mathbf{V}_l^\dagger\mathbf{Q}_l^\dagger\mathbf{H}_l\mathbf{x} + \boldsymbol{\eta} + \boldsymbol{\delta}_l \\
    &= \mathbf{G}_l\mathbf{x} + \boldsymbol{\eta} + \boldsymbol{\delta}_l
\end{aligned}
\end{equation}
The sum capacity is then given by
\begin{equation}
\label{eqn:sum_cap}
    \mathcal{C}_{\mathrm{sum}} = \log_2\det\Big(\mathbf{I}_K + \rho\sum_{l=1}^L\mathbf{G}_l^\dagger\big(\mathbf{\Phi}_l + \mathbf{I}_N\big)^{-1}\mathbf{G}_l\Big).
\end{equation}

\subsection*{Linear Symbol Detection}
The processing required to achieve (\ref{eqn:sum_cap}) is often prohibitively high, particularly for large networks. Lower complexity linear methods, such as linear minimum mean square error (LMMSE) symbol detection, are known to give near-optimal performance in systems with an excess of antennas. 
The LMMSE symbol estimate is given by
\begin{equation}
    \hat{\mathbf{x}} = \sum_{l=1}^L\mathbf{W}_l\tilde{\mathbf{z}}_l,
\end{equation}
with
\begin{equation}
    \mathbf{W}_l = \rho\Big(\mathbf{I}_K + \rho\sum_{i=1}^L\mathbf{G}_i^\dagger\big(\mathbf{\Phi}_i + \mathbf{I}_N\big)^{-1}\mathbf{G}_i\Big)^{-1}\mathbf{G}_l^\dagger\big(\mathbf{\Phi}_l + \mathbf{I}_N\big)^{-1}.
\end{equation}
The capacity of user $k$ under LMMSE symbol detection is
\begin{equation}
    \mathcal{C}_k = \log_2\big(1 + \mathrm{SQINR}_k\big),
\end{equation}
where $\mathrm{SQINR}_k$ is the signal-to-quantisation-plus-interference-plus-noise ratio of user $k$
\begin{equation}
\label{eqn:SQINR}
    \mathrm{SQINR}_k = \dfrac{1}{\Big[\Big(\mathbf{I}_K + \rho\sum_{l=1}^L\mathbf{G}_l^\dagger\big(\mathbf{\Phi}_l + \mathbf{I}_N\big)^{-1}\mathbf{G}_l\Big)^{-1}\Big]_{k,k}} - 1.
\end{equation}

\section{Imperfect CSI}
\label{sec:CSI}
Assuming MMSE channel estimation, the dimension reduction method can be readily adapted for the case of imperfect CSI at the receivers. The channel may be written,
\begin{equation}
    \mathbf{H}_l = \hat{\mathbf{H}}_l + \mathbf{E}_l
\end{equation}
where $\hat{\mathbf{H}}_l$ is the channel estimate and $\mathbf{E}_l$ the channel estimate error. For a given channel realisation and estimate, $\mathbf{E}_l$ is fixed (and unknown), but for random channels can be treated as a random variable, which by the orthogonality principle of MMSE estimation is uncorrelated with the channel estimate. Using the method outlined in \cite{1193803} the signal through the unknown channel may be absorbed into an uncorrelated equivalent noise term, $\boldsymbol{\omega}_l \sim \mathcal{CN}(0,\boldsymbol{\Omega}_l)$, 
\begin{equation}
    \mathbf{y}_l = \hat{\mathbf{H}}_l\mathbf{x} + \boldsymbol{\omega}_l,
\end{equation}
where 
\begin{equation}
    \boldsymbol{\Omega}_l = \mathbf{I}_M + \rho\sum_{k=1}^K\mathbf{C}_{l,k}
\end{equation}
with $\mathbf{C}_{l,k}$ the channel estimation error covariance for the channel between user $k$ and receiver $l$. For a given channel estimate, a transform may be applied to $\mathbf{y}_l$ to whiten this equivalent noise
\begin{equation}
\begin{aligned}
    \check{\mathbf{y}}_l &= \boldsymbol{\Omega}_l^{-1/2}\mathbf{y}_l \\
    &= \check{\mathbf{H}}_l\mathbf{x} + \check{\boldsymbol{\omega}}_l
\end{aligned}
\end{equation}
where $\check{\mathbf{H}}_l = \boldsymbol{\Omega}_l^{-1/2}\hat{\mathbf{H}}_l$, so that $\check{\boldsymbol{\omega}}_l\sim \mathcal{CN}(0,\mathbf{I}_M)$. 
The MF-GS dimension reduction method can be performed on the whitened signal, $\check{\mathbf{y}}_l$, using equivalent whitened channel vectors, $\check{\mathbf{h}}_{l,k} = \boldsymbol{\Omega}_l^{-1/2}\hat{\mathbf{h}}_{l,k}$.

Transform coding compression is then applied as described in Section \ref{sec:comp}, with decorrelating transform and rate allocation calculated using the eigenvectors and eigenvalues of $\mathbf{Q}_l^\dagger\check{\mathbf{H}}_l\check{\mathbf{H}}_l^\dagger\mathbf{Q}_l$. However, since for a given channel realisation $\mathbf{E}_l$ is unknown, the variance of the scalars being quantised are also not perfectly known. For analytical tractability, here we assume that these variances \textit{are} perfectly known, and accordingly calculate the quantisation noise covariance as
\begin{equation}
    \big[\mathbf{\Phi}_l\big]_{i,i} = \frac{\big( \mathbf{V}_l^\dagger\mathbf{Q}_l^\dagger\boldsymbol{\Omega}_l^{-1/2}\big(\rho\mathbf{H}_l\mathbf{H}_l^\dagger + \mathbf{I}_M\big)\boldsymbol{\Omega}_l^{-1/2}\mathbf{Q}_l\mathbf{V}_l\big)_{i,i}}{2^{r_{l,i}} - 1}.
\end{equation}
This is reasonable since it is known that for quantisers with a small mismatch in input variance the performance loss is small \cite{1092819}. 

By the reasoning in \cite{1193803} the mean sum capacity can then be lower bounded
\begin{equation}
    \mathbb{E}\big[\mathcal{C}_{\mathrm{sum}}^{\scriptscriptstyle\mathrm{CSI}}\big] \geq \mathbb{E}\Big[\log_2\det\Big(\mathbf{I}_K + \rho\sum_{l=1}^L\hat{\mathbf{G}}_l^\dagger\big(\mathbf{\Phi}_l + \mathbf{I}_N\big)^{-1}\hat{\mathbf{G}}_l\Big)\Big],
\end{equation}
where 
\begin{equation}
    \hat{\mathbf{G}}_l = \mathbf{V}_l^\dagger\mathbf{Q}_l^\dagger\boldsymbol{\Omega}_l^{-1/2}\hat{\mathbf{H}}_l.
\end{equation}
Capacity bounds under linear detection are similarly found by replacing $\mathbf{G}_l$ with $\hat{\mathbf{G}}_l$ in (\ref{eqn:SQINR}).

\section{Practical Considerations}
\label{sec:practical}
\subsection{Signalling Overheads}
Whilst a full treatment of the signalling overheads associated with this scheme is beyond the scope of this paper, we note that:
\begin{itemize}
    \item the CP requires the full channel matrices $\mathbf{H}_l$ for the MF-GS algorithm ($KM$ values per receiver).
    \item the receivers require the indices of the selected MF vectors, $\mathcal{S}_l$, for signal compression ($N$ indices per receiver)
    \item the CP requires reduced channel matrices $\mathbf{G}_l$ for signal decompression and symbol detection ($KN$ values per receiver, or calculated from $\mathbf{H}_l$).
\end{itemize}
Assuming all CSI is initially obtained at the receivers, the signalling overheads between receivers and CP are dominated by the full CSI required by the CP for MF vector selection. Since we can expect the MF vector selections to be significantly influenced by the large scale fading characteristics of the channel, one potential way to reduce signalling overheads is to fix the $\mathcal{S}_l$ for a number of coherence blocks between which only the channel fast fading changes. The reduced channel matrices, $\mathbf{G}_l$, can then be updated at the receivers at each coherence interval using only local CSI, and transferred to the CP for signal decompression and detection, reducing CSI overheads by a factor $M/N$.

\subsection{Fixed-rate Scalar quantisation}
The analysis in Section \ref{sec:comp} assumes the use of optimal Gaussian scalar compression, requiring long block lengths and complex encoders and decoders. Fixed rate Lloyd-Max scalar quantiser achieves the same quantisation noise using an additional 1.4 bits per scalar \cite{952802}, but with unit block length, and represents an attractive alternative for practical implementation.

\section{Numerical Results}
\label{sec:numerical}
Figure shows the rate-capacity curves for the reduced compression scheme for different signal dimensions, $N$. For comparison, the cut set bound,
\begin{equation}
    \mathcal{C}_{\mathrm{sum}} \leq \min\Big(\mathcal{R}L,\mathcal{I}\big(\mathbf{y}_1,\ldots,\mathbf{y}_L;\mathbf{x}\big)\Big),
\end{equation}
is shown, which represents an upper bound for all compression schemes. Using $N = 2$ gives the highest capacity in the rate limited region due to the lower quantisation noise, and operates close to the cut-set bound. At higher fronthaul rates, the capacity is limited due to the reduced dimension, and $N$ must be increased to increase capacity. 
\begin{figure}[h]
    \centering
    \includegraphics[width=0.95\linewidth]{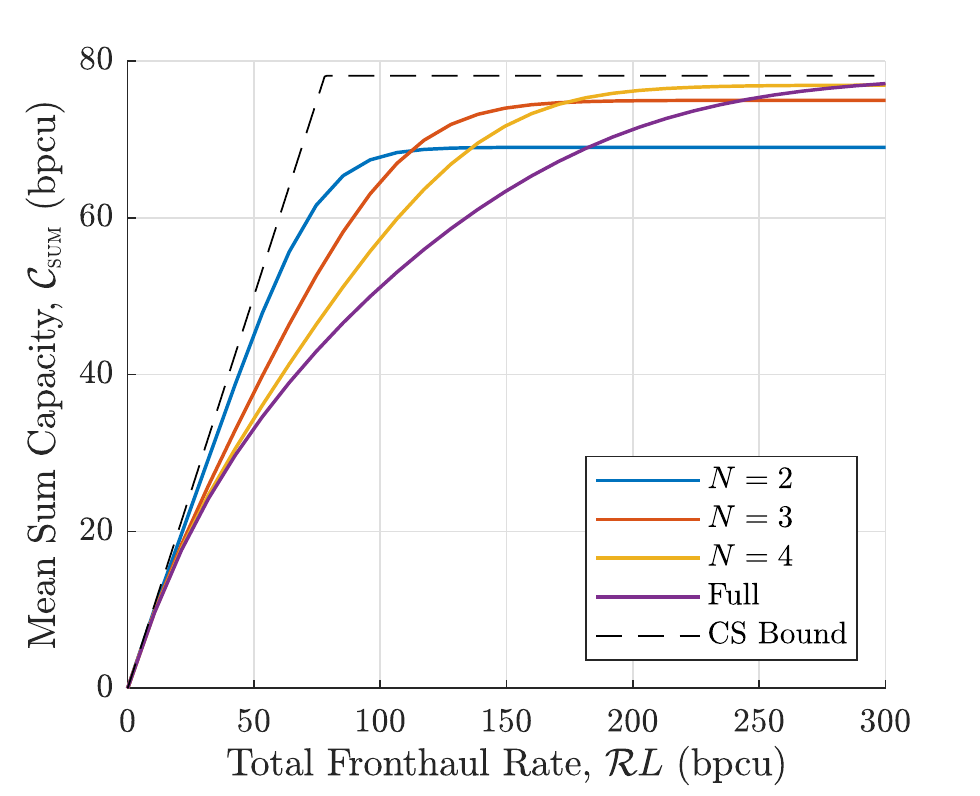}
    \caption{Rate-capacity performance for varying signal dimensions with $K=8$, $L=4$, $M = 8$, $\rho = 15$ dB.}
    \label{fig:Ncurves}
\end{figure}

\subsection{Sum Capacity}
\begin{figure}[h]
    \centering
    \includegraphics[width=0.95\linewidth]{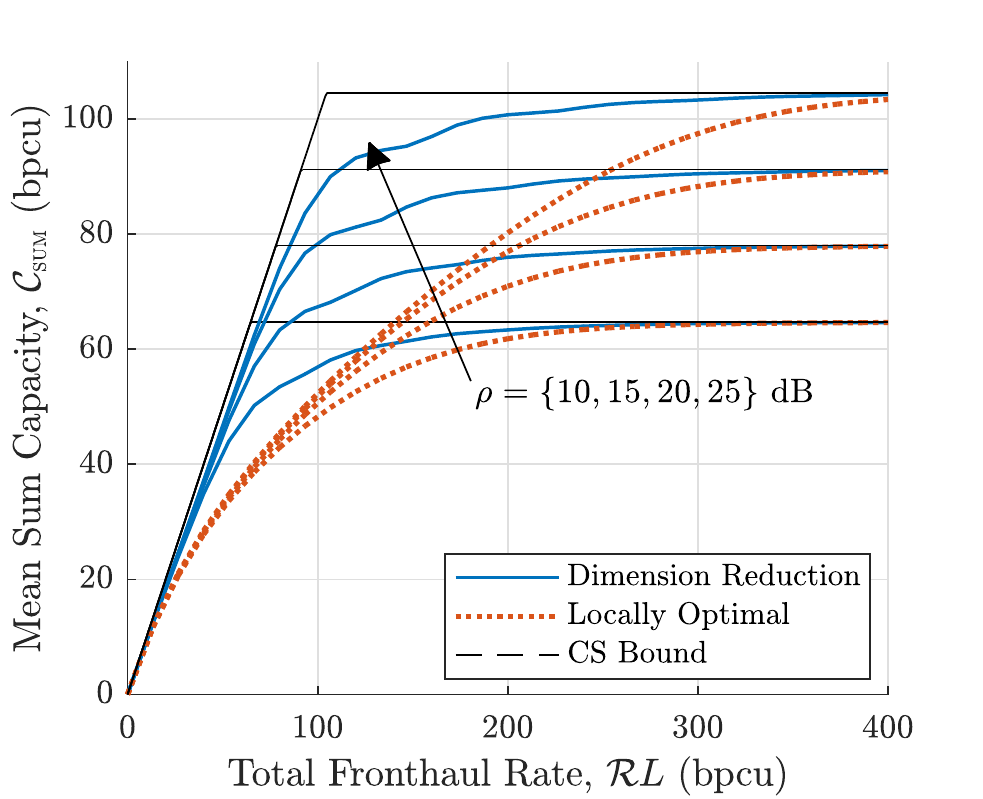}
    \caption{Rate-capacity performance for varying SNRs with $K=8$, $L=4$, $M = 8$.}
    \label{fig:rhocurves}
\end{figure}

Figure \ref{fig:rhocurves} shows the overall rate-capacity performance of the scheme, where for each value of $\mathcal{R}$, $N$ is chosen to maximise sum capacity. In practice this involves computing the equivalent channels for different values of $N$, which is simplified by noting that the MF-GS algorithm selects the same first $n'$ MF vectors for any $N\geq n'$, and for a given $\mathcal{R}$ we need only evaluate for a small range of $N$. 

We see that the scheme significantly outperforms standard local signal compression at all rates, and operates close to the cut-set bound in the rate limited region. The relative performance improvement of the scheme at high SNR can be understood with reference to Figure \ref{fig:dimredprop}, since at high SNR a small value of $N$ is able to capture an increased proportion of the total capacity.

\subsection{User Rates}
\begin{figure}[h]
    \centering
    \includegraphics[width=0.95\linewidth]{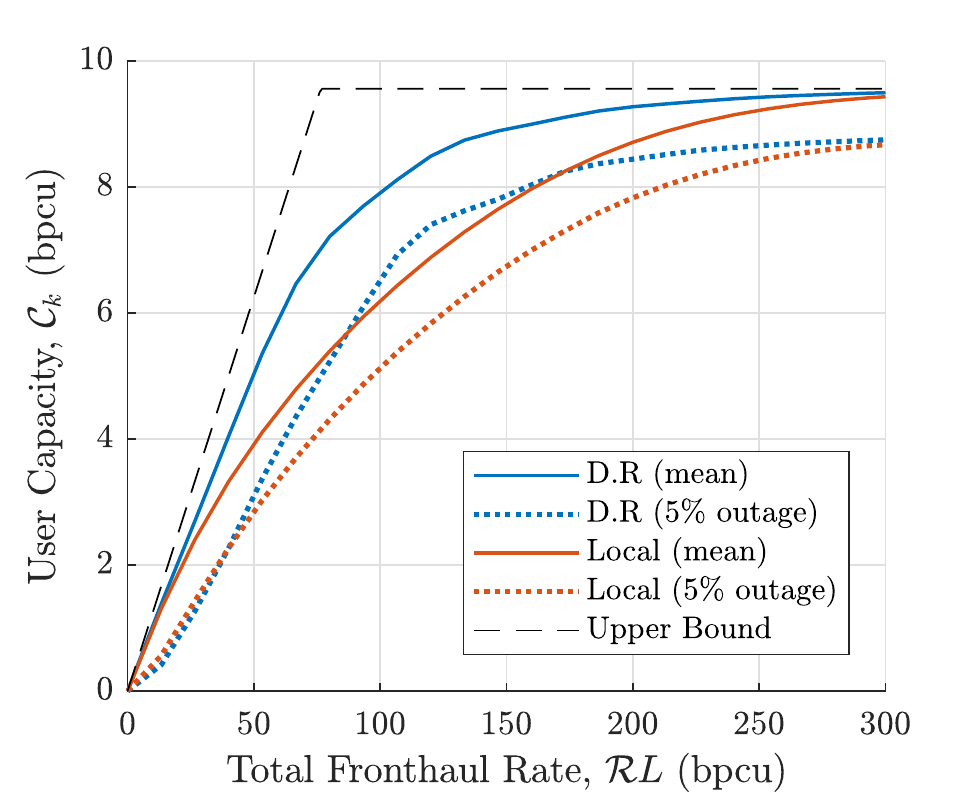}
    \caption{User mean and 5\% outage capacity with $K=8$, $L=4$, $M = 8$, $\rho = 15$ dB.}
    \label{fig:usercap}
\end{figure}

Figure \ref{fig:usercap} shows the mean and 5\% outage user capacities under dimension reduction compression. We see that the scheme offers a significant gain in both mean and outage capacity compared to local compression, for example an improvement of around 1.5 bpcu per user is achieved  at $\mathcal{R}L = 100$ bpcu.

\subsection{Imperfect CSI}
\begin{figure}[h]
    \centering
    \includegraphics[width=0.95\linewidth]{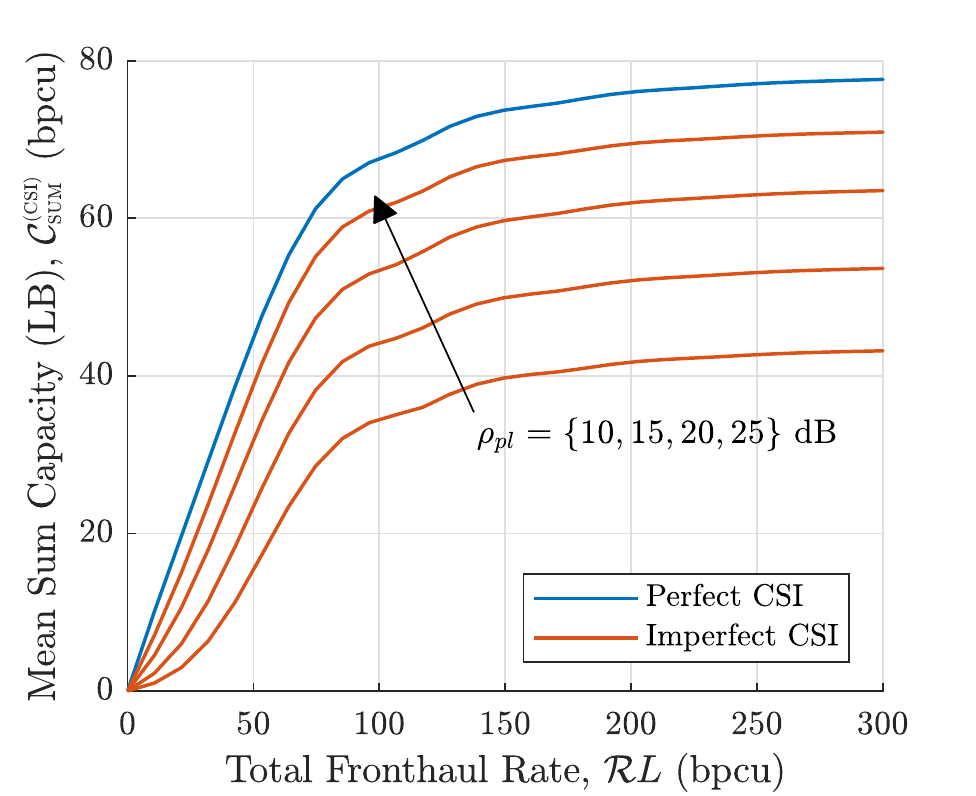}
    \caption{Rate-capacity lower bound for varying pilot SNR with $K=8$, $L=4$, $M = 8$, $\rho = 15$ dB.}
    \label{fig:CSIcurves}
\end{figure}

Figure \ref{fig:CSIcurves} shows the lower bound on sum capacity when MMSE channel estimation is performed using orthogonal uplink pilots with signal-to-noise ratio $\rho_{pl}$. When the CSI is good, the rate-capacity curve shows a similar shape to the perfect CSI case. With lower quality CSI, a fronthaul rate penalty is incurred due to the increased proportion of channel estimation error noise in the quantised signal.

\section{Conclusion}
In this work we have outlined a signal compression scheme for fronthaul-constrained distributed MIMO systems, based on applying dimension reduction prior to signal quantisation. Numerical examples demonstrate that the proposed dimension reduction algorithm is able to significantly reduce the number of signal components required at each receiver, and therefore significantly increase the rate-capacity performance of the scheme relative to local compression schemes - operating close to the cut-set capacity bound when the signal dimension is small. We further show that the scheme can be readily adapted for the case of imperfect CSI, and provide some practical suggestions for ways in which the signalling overheads and complexity of the quantisers can be reduced for implementation in practical systems.

\section*{Acknowledgement}
The work was supported by the Engineering and Physical Sciences Research Council grant number EP/I028153/1 and Toshiba Research Europe Limited.

\bibliography{arxiv.bib} 

\newcommand{\noop}[1]{}
\begin{thebibliography}{10}

\bibitem{brubaker2016emerging}
D.~Brubaker, W.~Qian, M.~Sussmann, Y.~Takafuji, M.~Akhter, and T.~Hiatt, ``The
  emerging need for fronthaul compression,'' {\em Altera Corporation White
  Paper}, pp.~1--12, 2016.

\bibitem{lombardi2018microwave}
R.~Lombardi, ``Microwave and millimetre-wave for 5g transport,'' {\em ETSI
  White Paper}, vol.~25, 2018.

\bibitem{7444125}
M.~{Peng}, Y.~{Sun}, X.~{Li}, Z.~{Mao}, and C.~{Wang}, ``Recent advances in
  cloud radio access networks: System architectures, key techniques, and open
  issues,'' {\em IEEE Communications Surveys Tutorials}, vol.~18,
  pp.~2282--2308, thirdquarter 2016.

\bibitem{6226311}
D.~{Samardzija}, J.~{Pastalan}, M.~{MacDonald}, S.~{Walker}, and
  R.~{Valenzuela}, ``Compressed transport of baseband signals in radio access
  networks,'' {\em IEEE Transactions on Wireless Communications}, vol.~11,
  pp.~3216--3225, Sep. 2012.

\bibitem{6824778}
Y.~{Zhou} and W.~{Yu}, ``Optimized backhaul compression for uplink cloud radio
  access network,'' {\em IEEE Journal on Selected Areas in Communications},
  vol.~32, pp.~1295--1307, June 2014.

\bibitem{7063645}
Y.~{Zhou} and W.~{Yu}, ``Optimized beamforming and backhaul compression for
  uplink mimo cloud radio access networks,'' in {\em 2014 IEEE Globecom
  Workshops (GC Wkshps)}, pp.~1493--1498, Dec 2014.

\bibitem{7134796}
L.~{Liu} and R.~{Zhang}, ``Optimized uplink transmission in multi-antenna c-ran
  with spatial compression and forward,'' {\em IEEE Transactions on Signal
  Processing}, vol.~63, pp.~5083--5095, Oct 2015.

\bibitem{8762078}
F.~{Wiffen}, M.~Z. {Bocus}, A.~{Doufexi}, and A.~{Nix}, ``Distributed mimo
  uplink capacity under transform coding fronthaul compression,'' in {\em ICC
  2019 - 2019 IEEE International Conference on Communications (ICC)}, pp.~1--6,
  May 2019.

\bibitem{8671721}
A.~{Liu}, X.~{Chen}, W.~{Yu}, V.~K.~N. {Lau}, and M.~{Zhao}, ``Two-timescale
  hybrid compression and forward for massive mimo aided c-ran,'' {\em IEEE
  Transactions on Signal Processing}, vol.~67, pp.~2484--2498, May 2019.

\bibitem{6920005}
B.~{Dai} and W.~{Yu}, ``Sparse beamforming and user-centric clustering for
  downlink cloud radio access network,'' {\em IEEE Access}, vol.~2,
  pp.~1326--1339, 2014.

\bibitem{7434656}
S.~{Sun}, T.~S. {Rappaport}, T.~A. {Thomas}, A.~{Ghosh}, H.~C. {Nguyen}, I.~Z.
  {Kovács}, I.~{Rodriguez}, O.~{Koymen}, and A.~{Partyka}, ``Investigation of
  prediction accuracy, sensitivity, and parameter stability of large-scale
  propagation path loss models for 5g wireless communications,'' {\em IEEE
  Transactions on Vehicular Technology}, vol.~65, pp.~2843--2860, May 2016.

\bibitem{6875354}
A.~{Winkelbauer}, S.~{Farthofer}, and G.~{Matz}, ``The rate-information
  trade-off for gaussian vector channels,'' in {\em 2014 IEEE International
  Symposium on Information Theory}, pp.~2849--2853, June 2014.

\bibitem{wiffen2020dimension}
F.~Wiffen, M.~Z. Bocus, W.~H. Chin, A.~Doufexi, and M.~Beach, ``Dimension
  reduction-based signal compression for uplink distributed mimo c-ran with
  limited fronthaul capacity,'' arXiv.org, 2020.

\bibitem{1193803}
B.~{Hassibi} and B.~M. {Hochwald}, ``How much training is needed in
  multiple-antenna wireless links?,'' {\em IEEE Transactions on Information
  Theory}, vol.~49, pp.~951--963, April 2003.

\bibitem{1092819}
R.~{Gray} and L.~{Davisson}, ``Quantizer mismatch,'' {\em IEEE Transactions on
  Communications}, vol.~23, pp.~439--443, April 1975.

\bibitem{952802}
V.~K. {Goyal}, ``Theoretical foundations of transform coding,'' {\em IEEE
  Signal Processing Magazine}, vol.~18, pp.~9--21, Sep. 2001.

\end{thebibliography}
\bibliographystyle{ieeetr}

\end{document}